# Designing the Mode solving of the photonic crystal fiber via BPM and Exploring the Single-Mode Properties

Mohammed DEBBAL[1], Mohamed CHIKH-BLED[2]

[1] Electronic Department, Abou-Bekr BELKAID University, Telecommunication Laboratory
Tlemcen, 13000, Algeria
*mohammed.debbal@mail.univ-tlemcen.dz*

[2] Electronic Department, Abou-Bekr BELKAID University, Telecommunication Laboratory
Tlemcen, 13000, Algeria
*mek_chikhbled@yahoo.fr*

**Abstract**

Microstructured optical fibers (MOFs) are one of the most exciting recent developments in fiber optics. A MOF usually consists of a hexagonal arrangement of air holes running down the length of a silica fiber surrounding a central core of solid silica or, in some cases air.
MOFs can exhibit a number of unique properties, including zero dispersion at visible wavelengths and low or high effective nonlinearity [3]–[17], By varying the size of the holes and their number and position, one can also design MOFs with carefully controlled dispersive and modal properties.
In this paper, we analyze and modeling the behavior of the photonic crystal fiber (PCF) by using in the first step a propagator method based on the BPM method.
With our BPM software, the electric field contour of the fundamental mode of PCF was demonstrated. We also used it to see the variation of the effective index; an effective index model confirms that such a fiber can be single mode for any wavelength.
It would make a study of photonic crystal fibers, and a study of the numerical simulation methods allow the simulation of optical properties and has modeled the propagation of light in this fiber type.
After that we use the V-parameter because it offers a simple way to design a photonic crystal fiber (PCF), by basing in a recent formulation of this parameter of a PCF, we provide numerically based empirical expression for this quantity only dependent on the two structural parameters, the air hole diameter and the hole-to-hole center spacing.

## 1. Introduction

Photonic Crystal Fibers (PCFs) [1]–[2], have been under intensive study for the past several years as they offer a number of unique and useful properties not achievable in standard silica glass fibers. PCFs fall into two basic categories.

The first one, an index-guiding PCF [3], [4], is usually formed by a central solid defect region surrounded by multiple air holes in a regular triangular lattice and confines light by total internal reflection like standard fibers.
The second one uses a perfect periodic structure exhibiting a photonic band-gap (PBG) effect at the operating wavelength to guide light in a low index core region, which is also called PBG fiber (PBGF) [5], [6].

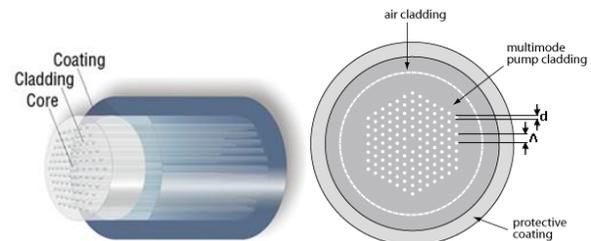

Fig. 1 : Structure of a Photonic Crystal Fiber with an air cladding.

A typical cross section of an index guided PCF is shown in Fig.1, The PCF consists of a triangular lattice of air holes where the core is defined by a "missing" air hole. The pitch is labeled $\Lambda$, and measures the period of the hole structure (the distance between the centers of neighboring air holes). The hole size is labeled d, and measures the diameter of the holes.

## 2. Beam Propagation Method (BPM)

The Beam Propagation Method (BPM) is a numerical modeling method to simulate the propagation of a wave in a guide of arbitrary geometry. It can predict from an incident field distribution within a structure. The main idea of this method is to divide a structure into "slices" elementary, spacing with ΔZ and then determine the scope of a given slice from the one before. However, the equations to solve are complex, which leads us to adopt certain approximations. [7]

## 3. Mode Solving via BPM

The earliest of these is referred to here as the correlation method, and was used to calculate modes and dispersion characteristics of multimode fibers.[8] More recently, a technique referred to as the imaginary distance BPM has been developed which is generally significantly faster.[9,10] It should be noted that the imaginary distance BPM technique is formally equivalent to many other iterative mode solving techniques;[11,12]

The results in [12], which can be duplicated via imaginary distance BPM, have shown excellent agreement with other published data.

Considering 2D propagation of a scalar field for simplicity, the incident field, $\emptyset_{in}(x$, can be expanded in the modes of the structure as

$$\emptyset_{in}(x) = \sum_m c_m \emptyset_m(x) \quad (1)$$

The summation should of course consist of a true summation over guided modes and integration over radiation modes, but for brevity the latter is not explicitly shown. Propagation through the structure can then be expressed as

$$\emptyset(x,z) = \sum_m c_m \emptyset_m(x) e^{iB_m Z} \quad (2)$$

In each BPM-based mode-solving technique, the propagating field obtained via BPM is conceptually equated with the above expression to determine how to extract mode information from the BPM results.
As the name implies, in the imaginary distance BPM the longitudinal coordinate z is replaced by z'=iz, so that propagation along this imaginary axis should follow

$$\emptyset(x,z') = \sum_m c_m \emptyset_m(x) e^{B_m Z'} \quad (3)$$

The propagation implied by the exponential term in Eq. 2 has become exponential growth in Eq. 3, with the growth rate of each mode being equal to its real propagation constant. The essential idea of the method is to launch an arbitrary field, say a Gaussian, and propagate the field through the structure along the imaginary axis. Since the fundamental mode (m=0) has by definition the highest propagation constant, its contribution to the field will have the highest growth rate and will dominate all other modes after a certain distance, leaving only the field pattern $\emptyset_0(x)$, The propagation constant can then be obtained by the following variational type expression:

$$\beta^2 = \frac{\int \emptyset^* \left(\frac{\partial^2 \emptyset}{\partial x^2} + k^2 \emptyset\right) dx}{\int \emptyset^* \emptyset \, dx} \quad (4)$$

Higher order modes can be obtained by using an orthogonalization procedure to subtract contributions from lower order modes while performing the propagation.[13] Issues such as optimal choice of launch field, reference wave number, and step size are discussed in [10,12]. Also, an additional correction is added which removes the error due to the fact that we have solved for the eigenvalues of the paraxial It is important to note that the imaginary distance BPM is not the same as the common technique of performing a standard propagation and waiting for the solution to reach steady state. The latter will only obtain the fundamental mode if the structure is single mode, and generally takes longer to converge. The imaginary distance BPM is closely related to the shifted inverse power method for finding eigenvalues and eigenvectors of a matrix.

In the correlation method, an arbitrary field is launched into the structure and propagated via normal BPM. During the propagation the following correlation function between the input field and the propagating field is computed:

$$p(z) = \int \emptyset^*_{in}(x) \, \emptyset(x,y) \, dx \quad (5)$$

Using Eq. 1 and Eq. 2, the correlation function can also be expressed as:

$$p(z) = \sum_m |c_m|^2 \, e^{iB_m Z} \quad (6)$$

From this expression one can see that a Fourier transform of the computed correlation function should have a spectrum with peaks at the modal propagation constants.
The corresponding modal fields can be obtained with a second propagation by beating the propagating field against the known propagation constants via:

$$\emptyset_m(x) = \frac{1}{L} \int_0^L \emptyset(x,z) \, e^{-i\beta_m Z} \quad (7)$$

Several corrections to the propagation constants can be made:

✓ A correction is made which accounts for the error introduced by solving the paraxial equation, and not the exact Helmholtz equation. Further details on the technique are found in [8].

✓ Second, the imaginary part of the propagation constant can be found by substituting the mode profile in the wave equation and solving for the propagation constant. This not only results in an imaginary value, but a corrected real value as well.

While the correlation method is generally slower than the imaginary distance BPM, it has the advantage that it is sometimes applicable to problems that are difficult or impossible for imaginary distance BPM, such as leaky or radiating modes.

## 4. Exploring the Single-Mode Condition

In order to explore the single-mode condition, we need to solve for the fundamental mode over a wide range of wavelengths. In a conventional fiber, the number of bound modes is governed by the $V$ number, which increases without limit as the wavelength decreases. The reference above shows that it is possible to define an effective $V$ number for a PCF that indicates reasonably accurately whether or not a fiber is single-mode:

$$V = \frac{2\pi}{\lambda} a \sqrt{n_{co}^2 - n_{cl}^2} = \sqrt{U^2 + W^2} \qquad (8)$$

with

$$U = \frac{2\pi}{\lambda} a \sqrt{n_{co}^2 - n_{eff}^2} \qquad (9)$$

$$W = \frac{2\pi}{\lambda} a \sqrt{n_{eff}^2 - n_{cl}^2} \qquad (10)$$

$\lambda$:  The wavelength
$a$:  is the core radius.
$n_{co}$:  is the core index.
$n_{cl}$:  is the cladding index.
$n_{eff}$:  is the effective index of fundamental guided mode.

$U$:  Normalized transversal phase constant.
$W$:  Normalized transversal attenuation constant.

V parameter determines the number of waveguide modes. That means the fiber is monomode only for the values of V (V<2.405) and if the value of V > 2.405, the fiber support more modes. An effective V value such as eq. (8) can be defined for the photonic crystal fiber eq. (11), for more detail see ref [4]:

$$V_{eff} = \frac{2\pi}{\lambda} \Lambda \sqrt{n_0^2 - n_{eff}^2} \qquad (11)$$

Here we use the pitch (center-to-center spacing) of the holes $\Lambda$.

The parameters defining $V_{eff}$ are all straightforward except for the effective cladding index $n_{eff}$.

We need a numerical method for obtaining the effective cladding index $n_{eff}$.

There are various software packages that can be used.

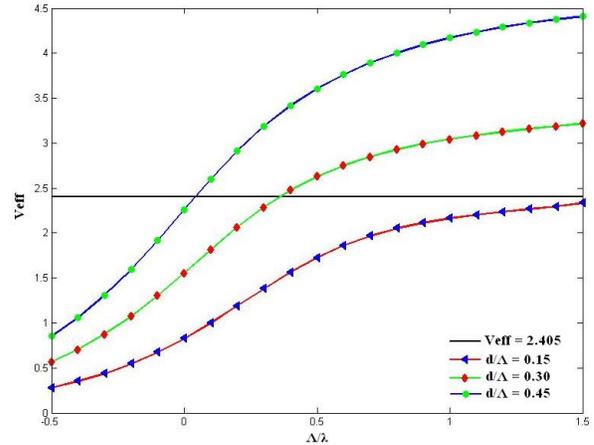

Fig. 2: Variation of Veff from Eq. 4 with d/Λ for various relative hole diameters: d/Λ= 0.15, d/Λ= 0.30 and d/Λ= 0.45, the dashed line indicate Veff = 2.405.

In Fig. 2, $V_{eff}$ is plotted as function of normalized frequency Λ/λ for values of d/Λ ranging from 0.15 to 0.45 in steps of 0.15.
The horizontal dashed line in plot indicates $V_{eff} = 2.405$ the condition for the fiber to be single-mode.

## 5. Dispersion features of the modes

The propagation in a PCF structure described in ref [8] has been simulated, the structure consists of an air-hole silica (n=1.45) index-guiding PCF characterized by a hexagonal distribution of air holes of radius (d = 0.6 μm) and pitch (Λ = 2.3 μm).

The mesh for the finite-difference computation has been generated using a spatial sampling step (Δx = Δy = Δz = 0.1 μm).

We used in our simulation both of the two structural, the first with a cylindrical hole, and the second with a square hole, as mention in fig 3.

We tried to modeling the phonic crystal fiber for the three (03) first modes, by using a BPM soft for a hexagonal structure of PCF with cylindrical air holes, and we calculate the Neff for these modes and we find as it mention in table 1.

After that the modal effective index of the modes has been computed by the BPM method, and the three electric filed components for both of the fundamental mode distribution and the three first modes are displayed in fig. 4 for λ=2.3.

Table 1: $N_{eff}$ of the three (03) first modes for hexagonal structure with cylindrical air holes

| Mode | $N_{eff}$ |
|---|---|
| 0 (fundamental mode) | 1.442815 |
| 1 | 1.448882 |
| 2 | 1.448851 |
| 3 | 1.446602 |

We did the same computing but this time with a hexagonal structure of PCF with square air holes, and we calculate the Neff for the fundamental and the three first modes, table 2.

Table 2: $N_{eff}$ of the three (03) first modes for hexagonal structure with square air holes

| Mode | $N_{eff}$ |
|---|---|
| 0 (fundamental mode) | 1.443218 |
| 1 | 1.448919 |
| 2 | 1.448882 |
| 3 | 1.446693 |

The three electric filed components for both of the fundamental mode distribution and the three first modes are displayed in fig. 5 for Λ=2.3.

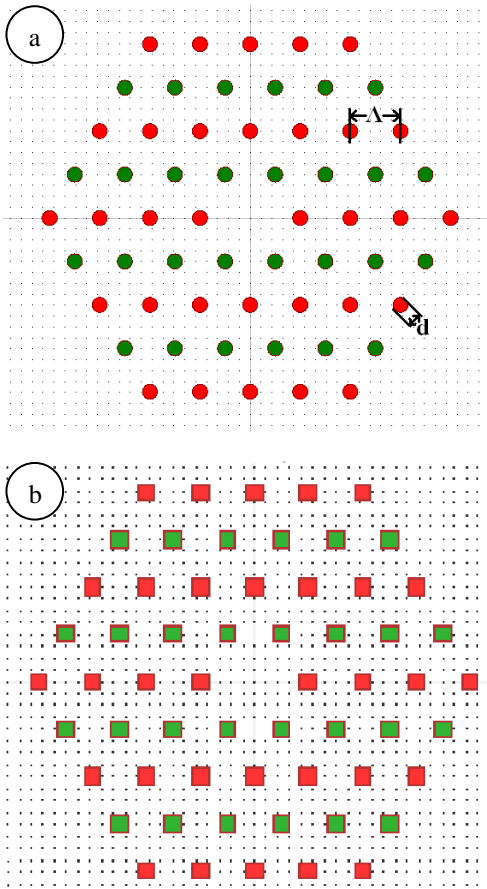

Fig. 3: the hexagonal structure of PCF with cylindrical air holes (a) and square air holes (b).

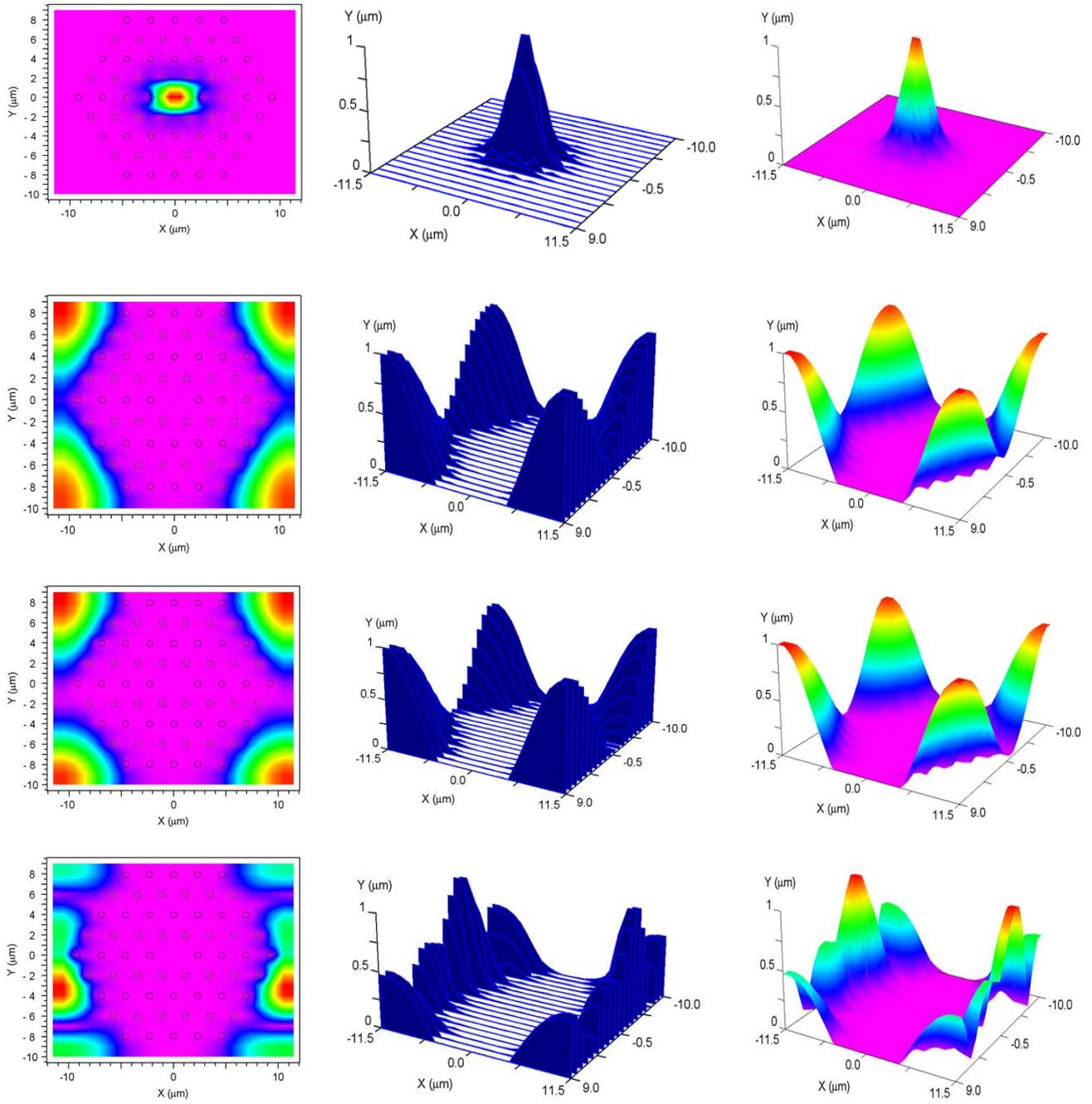

Fig. 4: x, y and z components of the electric field of the fundamental mode and three first modes of a PCF with cylindrical air holes for Λ=2.3.

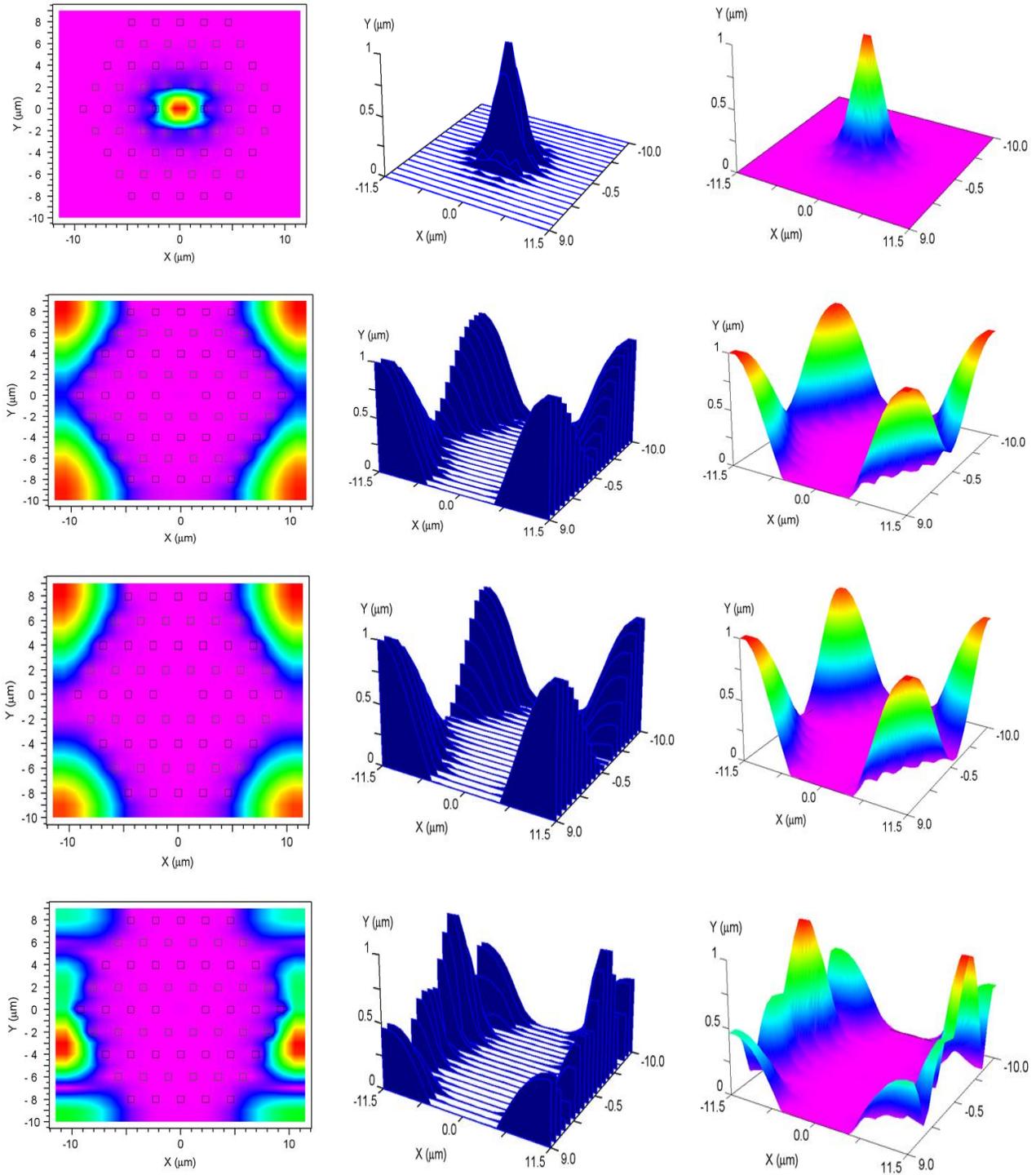

Figure 5: x, y and z components of the electric field of the fundamental mode and three first modes of a PCF with square air hoels for Λ =2.3.

## 6. The analytical Method

Nielsen and Mortensen are proved as [15] that V parameter can approximate with a function of type Eq. 5 depending on the wavelength λ, and the structural parameters d and Λ.

$$V\left(\frac{\lambda}{\Lambda}, \frac{d}{\Lambda}\right) = \frac{A\left(\frac{d}{\Lambda}\right)}{B\left(\frac{d}{\Lambda}\right) \times exp\left[C\left(\frac{d}{\Lambda}\right) \times \frac{\lambda}{\Lambda}\right] + 1} \quad (12)$$

In Fig. 6 we show V as function of λ/Λ for d/Λ ranging from 0.20 to 0.80 in steps of 0.05.

The parameter A, B and C are depend on d/λ only, and are describing by the following expressions:

$$A\left(\frac{d}{\Lambda}\right) = \frac{d}{\lambda} + 0.457 + \frac{3.405 \times \frac{d}{\Lambda}}{0.904 - \frac{d}{\Lambda}} \quad (13)$$

$$B\left(\frac{d}{\Lambda}\right) = 0.200 \times \frac{d}{\lambda} + 0.100 + 0.027 \times \left(1.045 - \frac{d}{\Lambda}\right)^{-2.8} \quad (14)$$

$$C\left(\frac{d}{\Lambda}\right) = 0.630 \times exp\left(\frac{0.755}{0.171 + \frac{d}{\Lambda}}\right) \quad (15)$$

The Eq. 12 constitutes the empirical expression for the V parameter in a PCF with λ/Λ and d/Λ being the only input parameter. For λ/Λ<2 and V>0.5 the expression gives values of V which deviates less than 3% from the correct values obtained from Eq.8 [14].

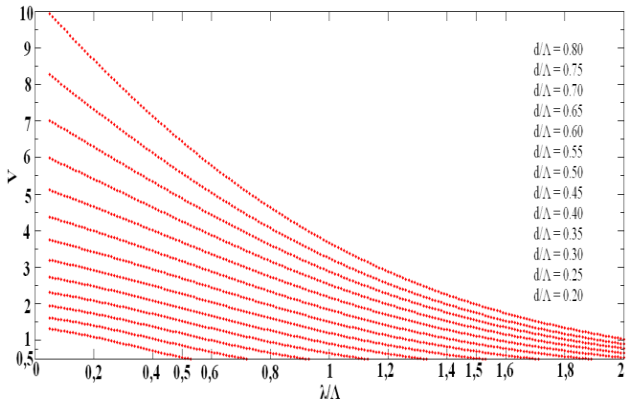

Fig. 6: V as a function of relative wavelength λ/Λ for d/Λ ranging from 0.20 to 0.80 in steps of 0.05.

## 7. Discussion and conclusion

Holey fiber technology is an attractive alternative to conventional technology, since it is possible to create extremely large mode area fibers that are single-mode over a broad wavelength range.

PCF with effective areas have been demonstrated that are effectively endlessly single-mode [16]. However, as with any fiber, the macro-bending losses place a fundamental upper limit on the mode sizes that are practical to use and are therefore an important consideration in the design of large mode area fibers.

Through this study we found that the normalized pitch Λ/λ and d/Λ are two factors that affect the effective index and the normalized frequency. Thanks to the simulation tool, we modeling and optimize the parameters of the microstructured fibers in order to design new components for optical telecommunications.

There are several issues to consider when designing a microstructured fiber; we have shown that the properties can be quantified via the V–parameter.

After that based on an extensive numeric calculation, we have established an empirical expression, which facilitates an easy evaluation of the V-parameter with the normalized wavelength and hole-size as the only input parameters for this experiment.

Photonic crystal fibers combine properties of 2D photonic crystals and classical fibers. Research on photonic crystal fibers is still very young and we may expect many new developments, more accurate and efficient methods for designing and optimization.